\documentstyle[aps]{revtex}

\input epsf

\begin{document}
\title{Depinning at the initial stage of the resistive \ transition \ in \ \
superconductors with a fractal cluster structure}
\author{Yuriy I. Kuzmin}
\address{Ioffe Physical Technical Institute of the Russian Academy of Sciences,\\
Polytechnicheskaya 26 St., Saint Petersburg 194021 Russia,\\
and State Electrotechnical University of Saint Petersburg,\\
Professor Popov 5 St., Saint Petersburg 197376 Russia\\
e-mail: yurk@mail.ioffe.ru; iourk@yandex.ru\\
tel.: +7 812 2479902; fax: +7 812 2471017}
\date{\today}
\maketitle

\begin{abstract}
Depinning of vortices in percolative superconductor containing fractal
clusters of a normal phase is considered. Transition of the superconductor
into a resistive state corresponds to the percolation transition from a
pinned vortex state to a resistive state when the vortices are free to move.
The motion of the magnetic flux transferred by these vortices gives rise to
the region of initial dissipation on current-voltage ({\it U}-{\it I})
characteristic. The influence of normal phase clusters on distinctive
features of {\it U}-{\it I} characteristics of percolative type-II
superconductors is considered. It is found that an increase in the fractal
dimension of the normal phase clusters causes the initial dissipation region
to broaden out. The reason of this effect is an increase in the density of
free vortices broken away from the pinning centers by the Lorentz force.
Dependencies of the free vortex density on the fractal dimension of the
normal phase cluster boundaries are obtained.
\end{abstract}

\bigskip

The dynamics of vortices in superconductors with fractal boundaries between
the normal and superconducting phases has recently received much attention
\cite{sur} -\cite{prb}. The study of their {\it U}-{\it I} characteristics
enable to get new information on the electromagnetic properties as well as
on the nature of a vortex state in such materials. New problems encountered
in this field are of interest in view of their importance for the
application of superconducting composites in electronics and power
engineering, especially, for superconducting wire fabrication.
Superconductors containing fractal clusters have specific magnetic and
transport properties \cite{pla3}, \cite{jltp}. The possibility to increase
the critical current through the enhancement of pinning by the fractal
normal phase clusters is of a particular interest \cite{tpl},\cite{pla2}.

This paper is devoted to an analysis of the initial region of the resistive
transition where the energy dissipation sets-in. Here the process of vortex
depinning gradually accrues resulting finally in the destruction of
superconducting state because of the thermo-magnetic instability. The
problem of initial dissipation in high-temperature superconductors (HTS's)
has been studied by many authors \cite{pre}, \cite{fu} -\cite{po}, \cite
{preSST}, \cite{preSPIE}. The residual resistance of bismuth-based HTS's [Bi$%
_{2}$Sr$_{2}$Ca$_{2}$Cu$_{3}$O$_{10+y}$ (BSCCO-2223) and Bi$_{2}$Sr$_{2}$CaCu%
$_{2}$O$_{10+y}$ (BSCCO-2212)] at small currents has been explained by
deterioration of grain boundaries, by initiation of micro-cracks which can
act as chains of serially connected weak links, as well as by the
grain-to-grain misalignment or by the degradation of grains themselves \cite
{fu}, \cite{su}. The ohmic behavior of {\it U}-{\it I} curves on the initial
stage of resistive transition in BSCCO-2223 and BSCCO-2212 has been
attributed to the local transfer of excess current into the normal metal
inclusions \cite{po}. The fractal regime in the initial stage of dissipation
has been observed in BSCCO-2223, YBa$_{2}$Cu$_{3}$O$_{7-x}$ (YBCO), and GdBa$%
_{2}$Cu$_{3}$O$_{7-x}$ \cite{pre}. The fractal nature of the normal phase
clusters in YBCO thin films has been found \cite{pla} and the effect of such
clusters on vortex dynamics has been analyzed \cite{prb} -\cite{jltp}, \cite
{pla2}.

Let us consider a superconductor containing inclusions of a normal phase,
which are out of contact with one another. We will suppose that the
characteristic sizes of these inclusions far exceed both the superconducting
coherence length and the penetration depth. A prototype of such a structure
is a superconducting wire.

The first generation HTS wires are fabricated following the powder-in-tube
technique (PIT). The metal tube is being filled with HTS powder, then the
thermal and deformation treatment is being carried out. The resulting
product is the wire (tape) consisting of one or more superconducting cores
armored by the normal metal sheath. The sheath endows the wire with the
necessary mechanical (flexibility, folding strength) and electrical (the
possibility to release an excessive power when the superconductivity will be
suddenly lost) properties. At the present time the best results are obtained
for the silver-sheathed bismuth-based composites, which are of practical
interest for energy transport and storage. In view of the PIT peculiarity
the first generation HTS wire has a highly inhomogeneous structure.
Superconducting core represents a dense conglomeration of BSCCO
micro-crystallites containing normal phase inclusions inside \cite{pas},
\cite{kov}. These inclusions primarily consist of a normal metal (silver) as
well as the fragments of different chemical composition, grain boundaries,
micro-cracks, and the domains of the reduced superconducting order
parameter. The volume content of the normal phase in the core is far below
the percolation threshold, so there is a percolative superconducting cluster
that carries the transport current.

The second generation HTS wires (coated conductors) have multi-layered film
structure consisting of the metal substrate (nickel-tungsten alloy), the
buffer oxide sub-layer, HTS layer (YBCO), and the protective cladding made
from the noble metal (silver). Superconducting layer, which carries the
transport current, has the texture preset by the oxide sub-layer. In the
superconducting layer there are clusters of columnar defects that can be
created during the film growth process as well as by the heavy ion
bombardment \cite{mez}. Such defects are similar in topology to the
vortices, therefore they suppress effectively the flux creep that makes
possible to get the critical current up to the depairing value \cite{ind}.

A passage of electric current through a superconductor is linked with the
vortex dynamics because the vortices are subjected to the Lorentz force
created by the current. In its turn, the motion of the magnetic flux
transferred by vortices induces an electric field that leads to the energy
losses. In HTS's the vortex motion is of special importance because of large
thermal fluctuations and small pinning energies \cite{blat}. Here we will
consider the simplified model of 1D pinning when a vortex filament is
trapped on the set of pinning centers \cite{blat2}. Superconductors
containing separated clusters of a normal phase provides for effective
pinning, because the magnetic flux is locked in these clusters, so the
vortices cannot leave them without crossing the superconducting space. These
clusters present the sets of normal phase inclusions, united by the common
trapped flux and surrounded by the superconducting phase \cite{pla}, \cite
{prb}. The magnetic flux can be created both by an external source (e.g., at
the magnetization in the field cooling regime) and by the transport current
(in the self-field regime). In the latter case the flux is concentrated
along irregular-shaped rings, which are deformed in such a way that the
normal phase clusters would be most captured. When the current is increased,
the vortices start to break away from the clusters of pinning force weaker
than the Lorentz force created by the transport current. As this takes
place, the vortices will first pass through the weak links, which connect
the normal phase clusters between themselves. In such a system depinning has
a percolative character \cite{ya}, \cite{zi}, \cite{zi2}, because unpinned
vortices move through the randomly generated channels created by weak links.
Weak links join the adjacent cells and enable the vortices to pass from one
cluster to another. Weak links form readily in HTS's due to the
intrinsically short coherence length \cite{son}. Depending on the specific
weak link configuration each normal phase cluster has its own current of
depinning, which contributes to the total statistical distribution of
critical currents.

One the other hand, weak links do not only provide for the percolation of
magnetic flux, but they also connect superconducting domains between
themselves, maintaining the electrical current percolation. As the transport
current is increased, the local currents flowing through ones or other weak
links begin to exceed the critical values, therefore some part of them
become resistive. Thus, the number of weak links involved in the
superconducting cluster is randomly reduced so the transition of a
superconductor into a resistive state corresponds to breaking of the
percolation through a superconducting cluster. The transport current acts as
a random generator that changes the relative fractions of conducting
components in classical percolative medium \cite{stau}, hence the resistive
transition can be treated as a current-induced critical phenomenon \cite{pre}%
.

The depinning current of each cluster is related to the cluster size,
because the larger cluster has more weak links over its boundary with the
surrounding superconducting space, and thus the smaller current of depinning
\cite{prb}. As a measure of the cluster size we will take the area of its
cross-section, and in the subsequent text we will call this value simply
``the cluster area``. An important feature of normal phase clusters is that
they can have fractal boundaries \cite{pla}, i. e. the perimeter of their
cross-section and the enclosed area obey the scaling law: $P^{1/D}\propto
A^{1/2}$, where $D$ is the fractal dimension of the cluster boundary \cite
{mandel}.

After the vortices start to move, a superconductor passes into a resistive
state. The voltage $U$, arising across the superconductor when the transport
current $I$ is passed through, can be expressed as a convolution integral,
in which the contributions from depinning currents of all the clusters are
taken into account:
\begin{equation}
U=R_{f}\int\limits_{0}^{I}\left( I-I^{\prime }\right) f\left( I^{\prime
}\right) dI^{\prime }  \label{volt1}
\end{equation}
where $R_{f}$ is the flux flow resistance, $f\left( I\right) $ is the
distribution function of the depinning currents.

In most practically important cases the distribution of the cluster areas
may be described by gamma distribution \cite{pla3} with the probability
density
\begin{equation}
w\left( a\right) =\frac{\left( g+1\right) ^{g+1}}{\Gamma \left( g+1\right) }%
a^{g}\exp \left( -\left( g+1\right) a\right)  \label{gamma2}
\end{equation}
where $\Gamma (\nu )$ is Euler gamma function, $a\equiv A/\overline{A}$ is
the dimensionless area of the cluster, $A$ is the area of the cross-section
of the cluster by the plane, transversally to which the vortices are moving,
$A_{0}>0$ and $g>-1$ are the parameters of gamma distribution that control
the mean area of the cluster $\overline{A}=\left( g+1\right) A_{0}$ and its
variance $\sigma
{2 \atop A}%
=\left( g+1\right) A%
{2 \atop 0}%
$. The mean dimensionless area of the cluster is equal to unity, whereas the
variance is determined by $g$-parameter only: $\sigma
{2 \atop a}%
=1/\left( g+1\right) $.

Gamma distribution of the cluster area of Eq.~(\ref{gamma2}) gives rise to
exponential-hyperbolic distribution of depinning currents \cite{pla3}
\begin{equation}
f(i)=\frac{2G^{g+1}}{D\Gamma (g+1)}i^{-\left( 2/D\right) (g+1)-1}\exp \left(
-Gi^{-2/D}\right)  \label{gamcur3}
\end{equation}
for which integral of Eq.~(\ref{volt1}) has the form:
\begin{equation}
u=\frac{r_{f}}{\Gamma \left( g+1\right) }\left( i\Gamma \left(
g+1,Gi^{-2/D}\right) -G^{D/2}\Gamma \left( g+1-\frac{D}{2},Gi^{-2/D}\right)
\right)  \label{uigam4}
\end{equation}
where $G\equiv \left( \theta ^{\theta }/\left( \theta ^{g+1}-\left(
D/2\right) \exp \left( \theta \right) \Gamma \left( g+1,\theta \right)
\right) \right) ^{2/D}$, $\theta \equiv g+1+D/2$, $i\equiv I/I_{c}$ is the
dimensionless electrical current normalized to the critical current of the
transition into a resistive state $I_{c}=\alpha \left( A_{0}G\right) ^{-D/2}
$, $\alpha $ is the form factor of the cluster, $D$ is the fractal dimension
of the cluster boundary, $\Gamma (\nu ,z)$ is the complementary incomplete
gamma function. The voltage across a superconductor $U$ and flux flow
resistance $R_{f}$ are related to the corresponding dimensionless quantities
$u$ and $r_{f}$ by the relationship: $U/R_{f}=I_{c}(u/r_{f})$.

In the simplest case of $g=0$, when gamma distribution of Eq.~(\ref{gamma2})
is reduced to the exponential one, $w(a)=exp(-a)$, the {\it U}-{\it I}
characteristics of Eq.~(\ref{uigam4}) can be written as
\begin{equation}
u=r_{f}\left( i\exp \left( -Ci^{-2/D}\right) -C^{D/2}\Gamma \left( 1-\frac{D}{%
2},Ci^{-2/D}\right) \right)  \label{uiexp5}
\end{equation}
where $C\equiv \left( \left( 2+D\right) /2\right) ^{2/D+1}$ is the constant
depending on the fractal dimension.

\begin{figure}[tbp]
\epsfbox{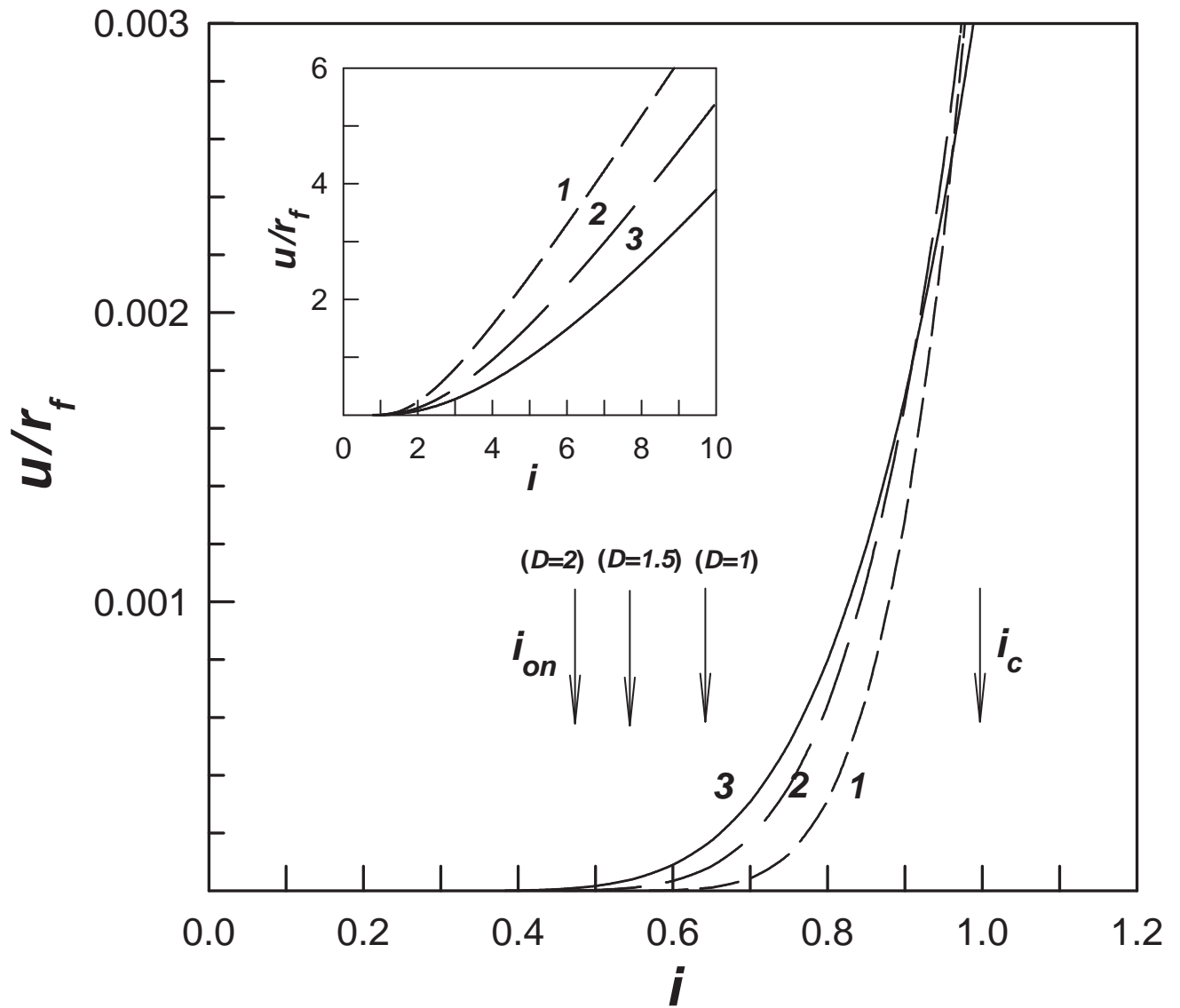}
\caption{The current-voltage characteristics
of a superconductor with fractal clusters of different fractal
dimensions $D=1$ (1), $D=1.5$ (2), and $D=2$ (3). The arrows
indicate the dissipation onset currents $i_{on}$ found on the
level of $10^{-5}u/r_{f}$ and the critical current $i_{c}$ of the
resistive transition.} \label{figure1}
\end{figure}

The corresponding {\it U}-{\it I} characteristics are shown in Fig.~\ref
{figure1}. The inset in this figure demonstrates that in the range of the
currents $i>1$ the fractality of the clusters reduces the voltage arising
from the magnetic flux motion. Meanwhile, the situation is quite different
in the neighborhood of the resistive transition below the critical current.
When $i<1$, the higher the fractal dimension of the normal phase cluster,
the larger is the voltage across a sample and the more stretched is the
region of initial dissipation in {\it U}-{\it I} characteristic. For further
consideration it is convenient to introduce the onset current $i_{on}$,
starting from which this region spreads away. The magnitude of this current
is set by the resolution of voltage measurement. In Fig.~\ref{figure1} the
arrows indicate the onset current values $i_{on}$ corresponding to the
resolution level of $10^{-5}u/r_{f}$. The inset in the same figure shows the 
dependence of the onset current $i_{on}$ on the fractal dimension. 
Figure~\ref{figure1} demonstrates that the value of the onset current decreases 
with increasing the fractal dimension, that is to say that the region of 
initial dissipation widens.

A significant difference in {\it U}-{\it I} characteristics before and after
the resistive transition is related to the dependence of the density of free
vortices on the fractal dimension for various transport currents. The free
vortex density determines the resistance, because the more vortices are free
to move the higher electric field is induced, and therefore, the greater is
the voltage at the same magnitude of a transport current. The density of
vortices broken away from pinning centers by the transport current i can be
found from the distribution of the depinning currents of Eq.~(\ref{gamcur3}%
),
\begin{equation}
n=\frac{B}{\Phi _{0}}\int\limits_{0}^{i}f\left( i^{\prime }\right)
di^{\prime }=\frac{B}{\Phi _{0}}F\left( i\right)  \label{vortex6}
\end{equation}
where $F(i)$ is the cumulative probability function of the depinning
currents, $B$ is the magnetic field, $\Phi _{0}\equiv hc/\left(
2e\right) $ is the magnetic flux quantum, $h$ is the Planck constant, $c$ is
the speed of light, and $e$ is the electron charge. The differential
resistance of a superconductor (which gives the slope of the {\it U-I}
characteristic) is proportional to the density of free vortices: $%
R_{d}=R_{f}(\Phi _{0}/B)n$.

For the exponential-hyperbolic critical current distribution of Eq.~(\ref
{gamcur3}), in the case of exponential distribution of the cluster areas $%
(g=0)$, the cumulative probability function has the form $F(i)=\exp
(-Ci^{-2/D})$. Therefore, the dependence of the free vortex density on the
fractal dimension can be written as
\begin{equation}
n\left( D\right) =\frac{B}{\Phi _{0}}\exp \left( -\left( \frac{2+D}{D}%
\right) ^{2/D+1}i^{-2/D}\right)  \label{expvortex7}
\end{equation}
In the special case of Euclidean clusters $(D=1)$, the formula of Eq.~(\ref
{expvortex7}) becomes: $n(D=1)=(B/\Phi _{0})\exp (-3.375/i^{2})$.

\begin{figure}[tbp]
\epsfbox{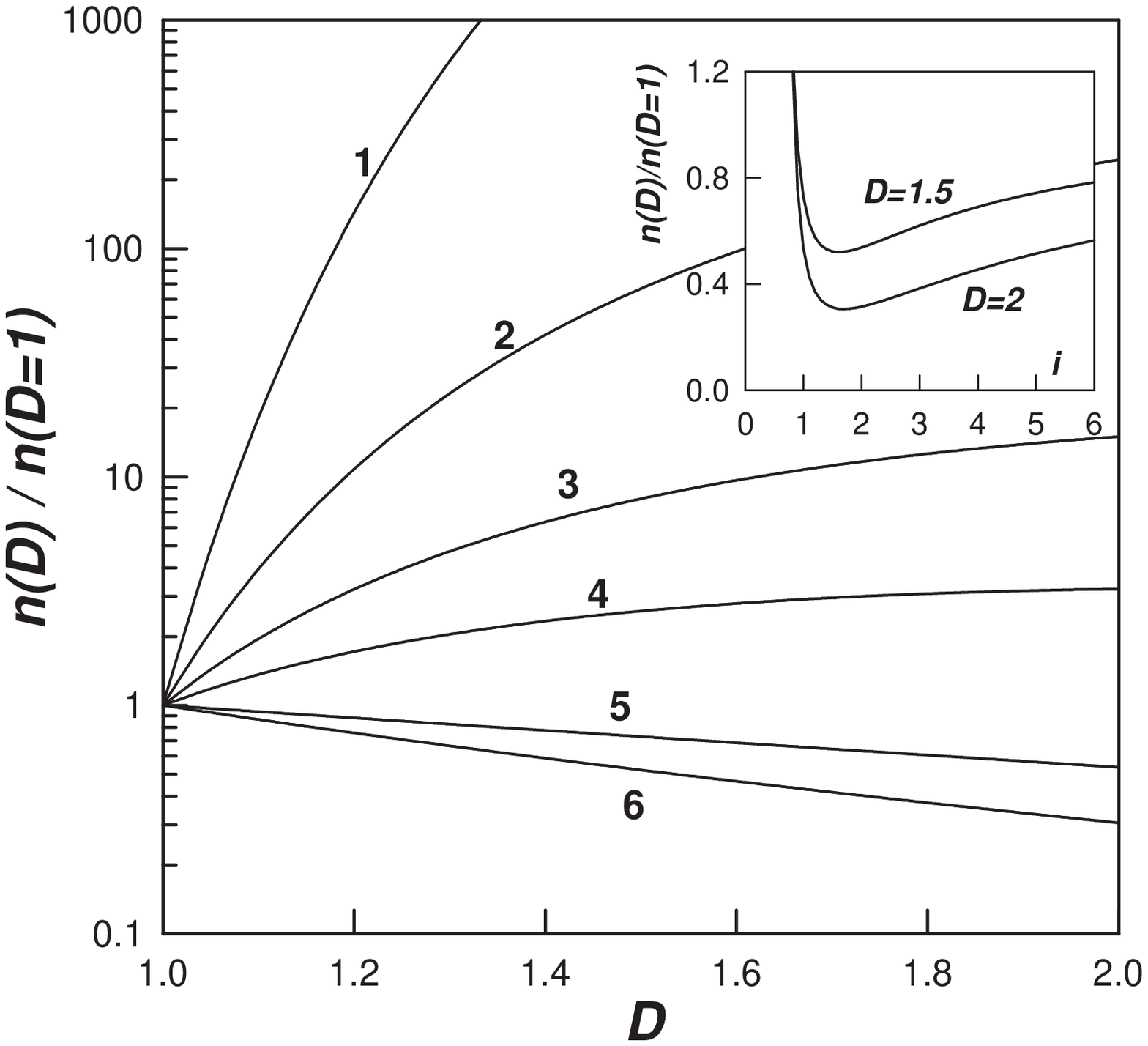}
\caption{Dependence of the free vortex
density on the fractal dimension of the normal phase clusters for
different values of a transport current $i=0.4$ (1), $i=0.5$ (2),
$i=0.6$ (3), $i=0.7$ (4), $i=1$ (5), $i=1.6875$ (6). The inset
shows the free vortex density versus current for two values of a
fractal dimensions $D=1.5$ and $D=2$.} \label{figure2}
\end{figure}

Figure \ref{figure2} demonstrates dependence of the relative density of free
vortices $n(D)/n(D=1)$ (relatively to the value for clusters with Euclidean
boundary) on the fractal dimension for different values of transport
currents. The vortices are broken away from pining centers mostly when $%
(i>1) $, that is to say, above the resistive transition. Here the free
vortex density decreases with increasing the fractal dimension. Such a
behavior can be explained by the fact that the critical current distribution
of Eq.~(\ref{gamcur3}) broadens out, moving towards greater magnitudes of
current as the fractal dimension increases. It means that more and more
clusters of high depinning current, which can trap the vortices best, are
being involved in the game. The smaller part of the vortices is free to move
the smaller the induced electric field. An important feature of the
superconductors with fractal clusters is that fractality of the cluster
boundary enhances pinning and, hence, a current-carrying capability of the
superconductor. The relative change in free vortex density depends on the
transport current (see inset in Fig.~\ref{figure2}) and in the limiting case
of the most fractal boundary $(D=2)$ reaches a minimum for $i=1.6875$ (curve
6 in Fig.~\ref{figure2} goes below others). That corresponds to the maximum
pinning gain and the minimum level of dissipation. As may be seen in Fig.~%
\ref{figure1}, the voltage across a sample carrying the same transport
current decreases with increasing fractal dimension.

In the range of transport currents below the resistive transition $(i<1)$,
the situation is different: resistance, as well as the free vortex density,
increases for the clusters of greater fractal dimension. Such a behavior is
related to the fact that the critical current distribution of Eq.~(\ref
{gamcur3}) broadens out, covering both high and small currents, as the fractal
dimension increases. In spite of sharp increase in relative density of free
vortices (Fig.~\ref{figure2}), the absolute value of vortex density in the
range of the currents involved is very small (much smaller than above the
resistive transition). So the vortex motion does not lead to destruction of
superconducting state yet, and the resistance remains very low. The low
density of vortices at small currents is related to the peculiarity of
exponential-hyperbolic distribution of Eq.~(\ref{gamcur3}). The cumulative
probability function $F\left( i\right) $ for this distribution is so
``flat`` at small currents, that all its derivatives are equal to zero at
the point of $i=0$: $d^{k}F(0)/di^{k}=0$ for any value of $k$. Even the
expansion of $F(i)$ into Taylor's series at this point tends to zero, rather
than to $F$. This behavior has a clear physical meaning. Indeed, so small a
transport current does not significantly affect the trapped magnetic flux
because there are scarcely any pinning centers of such small critical
currents, so that nearly all the vortices are still pinned. Significant
breaking of the vortices away begins only after the resistive transition when
$i>1$.

For any hard superconductor (of type-II, with pinning centers) the
dissipation in the resistive state does not mean the destruction of phase
coherence yet. Some dissipation always accompanies any motion of a magnetic
flux that can happen in a hard superconductor even at low transport current.
The superconducting state collapses only when a growth of dissipation
becomes avalanche-like as a result of thermo-magnetic instability.

Thus, the fractal properties of the normal phase clusters significantly
affect the vortex dynamics in superconductors. The {\it U}-{\it I}
characteristics of superconductors with a fractal cluster structure have two
distinctive regions - before and after the resistive transition. Each of
them has different dependence of free vortex density of on the fractal
dimension of normal phase clusters boundaries. After the resistive
transition the fractality of the clusters suppresses the vortex depinning,
thus increasing the current-carrying capacity of the superconductor.

\begin{center}
${\bf Acknowledgements\smallskip }$
\end{center}

This work is supported by the Russian Foundation for Basic Researches (Grant
No 02-02-17667).

\newpage

\begin{references}
\bibitem{sur}  R. Surdeanu, R. J. Wijngaarden, B. Dam, J. Rector, R.
Griessen, C. Rossel, Z. F. Ren, and J. H. Wang, Phys. Rev. B {\bf 58}, 12467
(1998).

\bibitem{pla}  Yu. I. Kuzmin, Phys. Lett. A {\bf 267}, 66 (2000).

\bibitem{prb}  Yu. I. Kuzmin, Phys. Rev. B {\bf 64}, 094519 (2001).

\bibitem{pla3}  Yu. I. Kuzmin, Phys. Lett. A {\bf 300}, 510 (2002).

\bibitem{jltp}  Yu. I. Kuzmin, J. Low Temp. Phys. {\bf 130}, 261 (2003).

\bibitem{tpl}  Yu. I. Kuzmin, Tech. Phys. Lett. {\bf 26}, 791 (2000).

\bibitem{pla2}  Yu. I. Kuzmin, Phys. Lett. A {\bf 281}, 39 (2001).

\bibitem{pre}  M. Prester, Phys. Rev. B {\bf 60}, 3100 (1999).

\bibitem{fu}  Y. Fukumoto, Q. Li, Y. L. Wang, M. Suenaga, and P. Haldar,
Appl. Phys. Lett. {\bf 66}, 1827 (1995).

\bibitem{su}  M. Suenaga, Y. Fukumoto, P. Haldar, T. R. Thurston, and U.
Wildgruber, Appl. Phys. Lett. {\bf 67}, 3025 (1995).

\bibitem{po}  M. Polak, W. Zhang, J. Parrell, X. Y. Cai, A. Polyanskii, E.
E. Hellstrom, D. C. Larbalestier, and M. Majoros, Supercond. Sci. Technol.
{\bf 10}, 769 (1997).

\bibitem{preSST}  M. Prester, Supercond. Sci. Technol. {\bf 11}, 333 (1998).

\bibitem{preSPIE}  M. Prester, P. Kovac, I. Husek, Proc. SPIE {\bf 3481}, 60
(1998).

\bibitem{pas}  A. E. Pashitski, A. Polyanskii, A. Gurevich, J. A. Parrell,
and D. C. Larbalestier, Physica C {\bf 246}, 133 (1995).

\bibitem{kov}  P. Kov\'{a}\v{c}, I. Hu\v{s}ek, W. Pachla, T. Meli\v{s}ek,
and V. Kliment, Supercond. Sci. Technol. {\bf 8}, 341 (1995).

\bibitem{mez}  E. Mezzetti, R. Gerbaldo, G. Ghigo, L. Gozzelino, B. Minetti,
C. Camerlingo, A. Monaco, G. Cuttone, and A. Rovelli, Phys. Rev. B {\bf 60},
7623 (1999).

\bibitem{ind}  M. V. Indenbom, M. Konczykowski, C. J. van der Beek, and F.
Holtzberg, Physica C {\bf 341--348}, 1251 (2000).

\bibitem{blat}  G. Blatter, M. V. Feigelman, V. B. Geshkenbein, A. I.
Larkin, and V. M. Vinokur, Rev. Mod. Phys. {\bf 66}, 1125 (1994).

\bibitem{blat2}  G. Blatter, V. B. Geshkenbein, and J. A. G. Koopmann, Phys.
Rev. Lett. {\bf 92}, 067009 (2004).

\bibitem{ya}  K. Yamafuji and T. Kiss, Physica C {\bf 258}, 197 (1996).

\bibitem{zi}  M. Ziese, Physica C {\bf 269}, 35 (1996).

\bibitem{zi2}  M. Ziese, Phys. Rev. B {\bf 53}, 12422 (1996).

\bibitem{son}  J. E. Sonier, R. F. Kiefl, J. H. Brewer, D. A. Bonn, S. R.
Dunsiger, W. N. Hardy, R. Liang, R. I. Miller, D. R. Noakes, and C. E.
Stronach, Phys. Rev. B {\bf 59}, R729 (1999).

\bibitem{stau}  D. Stauffer, Phys. Rep. {\bf 54}, 2 (1979).

\bibitem{mandel}  B. B. Mandelbrot, {\it The Fractal Geometry of Nature}
(Freeman, San Francisco, 1982).

\newpage
\end{references}
\end{document}